\documentclass[showpacs,preprintnumbers,amsmath,amssymb, prd]{revtex4}
\usepackage{amsmath,amssymb,graphicx}
\usepackage{epsf}
\usepackage{epstopdf}
\usepackage{color}
\usepackage {amssymb}
\newcommand{\nc}{\newcommand}
\nc{\ba}{\begin{eqnarray}}
\nc{\ea}{\end{eqnarray}}
\newcommand\be{\begin{equation}}
\newcommand\ee{\end{equation}}

\newcommand{\calR}{{\cal{R}}}

\nc{\x}{{\bf{x}}}
\nc{\calP}{{\cal P}}
\nc{\calF}{{\cal F}}

\begin{document}

\author{Mohammad Hossein Namjoo}
\email{mh.namjoo-AT-ipm.ir}

\address{School of Physics, Institute for Research in
Fundamental Sciences (IPM),
P.~O.~Box 19395-5531,
Tehran, Iran}

\author{Shant Baghram}
\email{baghram-AT-ipm.ir}
\address{School of Astronomy, Institute for Research in
Fundamental Sciences (IPM),
P.~O.~Box 19395-5531,
Tehran, Iran}

\author{Hassan Firouzjahi}
\email{firouz-AT-ipm.ir}

\address{School of Astronomy, Institute for Research in
Fundamental Sciences (IPM),
P.~O.~Box 19395-5531,
Tehran, Iran}

\vskip 1cm

\title{Hemispherical Asymmetry and Local non-Gaussianity: a Consistency Condition}

\begin{abstract}
In this paper we provide a consistency relation between the amplitude of the hemispherical bipolar asymmetry, $A$,  and the amplitude of  the primordial non-Gaussianity in the squeezed limit, $f_{NL}$, as $|A|\lesssim 10^{-1}f_{NL}$. We demonstrate that this consistency condition is at work for any model of inflation in which the curvature perturbations is sourced by a single light field with the Bunch-Davies initial condition,
irrespective of the number of inflation fields which contribute to the background inflationary expansion.  As a non-trivial example, we show that observable hemispherical asymmetry can be generated in single field non-attractor inflationary models. We also study hemispherical asymmetry generated   in the models of multiple fields inflation. We show that $A$ is controlled by the weighted sum of  non-Gaussianity contribution from each field.  In particular we show that observable hemispherical asymmetry can be generated  in models where inhomogeneities are generated from a light scalar field modulating  the surface of end of inflation.

\end{abstract}

\maketitle

\section{Introduction}
\label{introduction}

Cosmology is in the era of precision. The release of the Planck data \cite{Ade:2013zuv} confirmed the validity of  the Standard Model of Cosmology, $\Lambda$CDM, with the initial conditions from inflationary paradigm \cite{Guth:1980zm} seeding the late time Large Scale Structures (LSS) of the Universe, which are nearly Gaussian, adiabatic and scale-invariant \cite{Ade:2013uln}. However, the Planck collaboration also reported a hemispherical asymmetry in Cosmic Microwave Background temperature (CMB) fluctuations  \cite{Ade:2013nlj} which was also observed in Wilkinson Microwave Anisotropy Probe (WMAP) data \cite{Eriksen:2003db}. The statistical significance of this finding may be under debate
\cite{Bennett:2010jb, Hanson:2010gu}. However, should the upcoming observations confirm this finding, then this may be viewed as a challenge for simple models of inflation which predict an isotropic Universe with all pre-inflationary history  washed out during the near exponential expansion of the background \cite{Dai:2013kfa,{Pullen:2007tu}}.

The observed asymmetry may indicate to a pre-inflationary physics. Historically  Grishchuk and Zel'dovich\cite{Grishchuk1978} first considered the temperature anisotropy from superhorizon perturbations, then Erickcek et al. \cite{Erickcek:2008sm}   showed that a super-horizon perturbation may introduce this asymmetry, the idea which also  mentioned in Gordon\cite{Gordon:2006ag}. This is because the perturbations produced by the inflaton field depend on the background value of the field. Therefore the long wavelength mode (i.e. a super-horizon mode) changes the perturbations and are potentially capable to introduce asymmetry in curvature perturbation power spectrum. The decomposition of modes into the long (super-horizon) and short (CMB) scale modes and their add up is very similar to the Peak- Background Splitting of the modes in LSS  \cite{Sheth:1999mn}, where, in analogy,  the long wavelength mode (horizon size mode) changes the amplitude of the matter perturbation on short
scales (structures' scales ``i.e. cluster of galaxies").

 Recently a very interesting proposal was put forward by Schmidt and Hui   \cite{Schmidt:2012ky}, that the modulation from the primordial non-Gaussianities (PNG) can produce the asymmetry by mixing the long and short wavelength modes \cite{Prunet:2004zy}, for a related idea see also \cite{Byrnes:2011ri}. This is an interesting observation which can yield a bridge between the predictions from PNG and observations.  However, it is worth to mention that the asymmetry generated from the modulation of the large amplitude long wavelength mode will introduce large angular temperature fluctuations, like quadrupole and octupole anisotropies on  CMB, so the observed uniformity of CMB map constrain these departures from isotropy \cite{Erickcek:2008jp}. These constrains  mainly come from the quadrupole,  $C_2$, and the octupole, $C_3$, moments of CMB \cite{Grishchuk1978, Erickcek:2008jp,Turner:1991dn, Zibin:2008fe}. This is because the primordial curvature perturbations are translated into the perturbations in gravitational potential ($\Phi$) which causes the temperature fluctuations by Sachs-Wolfe (SW) effect ($\Delta T/T\simeq \Phi/3$)  \cite{Sachs:1967er}.  This gravitational potential modulation can not generate a monopole and a dipolar in CMB power spectrum. This is because they are canceled   by the angle averaged CMB power spectrum and the intrinsic dipolar term introduced from the peculiar velocity respectively \cite{Erickcek:2008jp,Turner:1991dn,Bruni:1993dx}. However, the quadrupole  and the octupole terms are distinguishable and also are well constrained by CMB observations \cite{Castro:2003bk}.
 
A phenomenological parameterizations of the dipolar asymmetry is defined via
\ba
\label{P-asym}
{\cal P}^{1/2}_{\cal R}(k, {\bf x}) = {\cal P}^{1/2^{iso}}_{{\cal R}}(k) (1+ A(k) {\bf \hat p.x}/x_n )
\ea
in which ${\cal P}_{\cal R}(k, {\bf x})$ is the curvature perturbations power spectrum, ${\cal P}^{1/2^{iso}}_{\cal R}$ is the   isotropic power spectrum, $A(k)$ is the amplitude of the bipolar asymmetry, $\bf \hat p$  is its direction and $x_n$ is the comoving distance to the surface of last scattering. The Planck data indicates $A =0.072 \pm 0.022$ on large angular scales, $\ell < 64$,  with the best fit for the anisotropy direction $(l,b)=(227,-27)$ \cite{Ade:2013nlj}.

In Erickcek et al. \cite{Erickcek:2008sm} it is shown that the maximum anisotropy generated in models of single clock inflation  is $A_{max}\simeq 0.01$ in order not to contradict with the observational constraints from the quadrupole and octupole constraints. This value of $A$ is almost an order of magnitude smaller than the central value of anisotropy amplitude reported by Planck.  One can easily check that in conventional models of single filed  inflation  $A \propto (n_s -1)$ in which $n_s$ is the curvature perturbation spectral index, which from the Planck observation, is obtained to be very close to unity. As a result, in simple single field inflation models, the smallness of $A$ is directly related to the
smallness of $n_s-1$. If one compares this situation with the Maldacena's consistency relation \cite{Maldacena:2002vr} for the amplitude of PNG in the squeezed limit, $f_{NL}$, in which $f_{NL} \sim n_s -1$, the similarity rings the bell. To make this similarity more pronounced,
it is observed that in curvaton models with large enough value of $f_{NL}$ an  observable value of $A$ can be generated \cite{Erickcek:2008sm, Lyth:2013vha}. These results tempt to indicate that in general the amplitude of  $A$ is controlled by $f_{NL}$. This was  directly demonstrated by Lyth in curvaton scenario in which he found $A \propto f_{NL}$.  We prove this conclusion in a model-independent way. We show that for all models of inflation in which the curvature perturbations have a single source with the Bunch-Davies initial condition  this consistency condition does hold.

There are additional constraints from LSS on the asymmetry from
power spectrum modulation. The SDSS sample of quasars sets a very strict constrain on $A$ in the scales of $k\sim 1 Mpc ^{-1}$  \cite{Hirata:2009ar}. Although the hemispherical asymmetry  shows up in very large scales,  $k< 0.035 Mpc^{-1}$, but this constraint implies that the power modulation must vanish very quickly in going from large to small scales.
As a result,  the two cosmological observations: a) CMB anisotropy constraints from quadrupole and octupole  and b) the constrains from the LSS (i.e. quasars)  on  the primordial anisotropic  power spectrum in small scales, restrict the space of the inflationary models based on super-horizon modulations of ${\cal P}_{\cal R}(k, {\bf x})$.
 On the other hand,  this may open up a new venue to investigate other cosmological models as the source of asymmetry,
like considering the modulations of the spectral index, the
modulation of the tensor perturbation amplitude, the modulation of the optical depth or spontaneous symmetry breaking \cite{Dai:2013kfa,Gordon:2005ai}. \\

{\bf Note added}: While this work was in its final stage, the paper \cite{Wang:2013lda} appeared which has some overlaps with our multiple fields analysis in Section \ref{multiple}.

\section{ Single Source Inflation and the Consistency Relation}
\label{single}

In this section we prove our consistency condition for models of inflation in which the curvature perturbation is generated by a single source. Of course there may be more than one inflaton fields at the background which can contribute to background expansion to solve the flatness and the horizon problem. But we assume that only one field is responsible for perturbations in order to generate the observed CMB fluctuations and the structure formation. With this definition, the standard curvaton model \cite{Lyth:2001nq, Linde:1996gt}
is also considered as a single source scenario in which the curvature perturbations originate entirely from the curvaton fluctuations. On the other hand, if both the inflaton field and the curvaton field contribute to perturbations, then it is a multiple source scenario which is studied in next section.

Having this said, one may wonder what is so crucial about the assumption of single source inflation. To see this, let us consider the curvature perturbation on comoving surface  $\cal R$.
With an appropriate choice of gauge transformation, one can always find a coordinate
system (comoving gauge)  in which the comoving velocity potential $\delta u$ vanishes. Furthermore, in  models in which perturbations are sourced by the single field $\phi$, one has $\delta u \propto \delta \phi$. Now, in the comoving gauge in which $\delta u= \delta \phi=0$,  the re-normalized scale factor $\tilde a (t, \x)$ on each Hubble patch (in the sense of separate Universe approach) is given by
 \ba
\label{a-renormalized}
\tilde a (t, \x) = a(t)  e^{ {\cal R} (t, \x)}
\ea
 in which $a(t)$ is the background scale factor.  As a result, the three-dimensional spatial metric on the surface of constant time is given by
 \ba
 \label{three-metric}
 ds_{(3)}^2 = a(t)^2 e^{ 2 {\cal R} (t, \x)} d \x^i d \x_i  \, .
 \ea
In this view, the effect of a large scale curvature perturbations will be a modulation of comoving distance $\x \rightarrow e^{ {\cal R} (t, \x)} \x$ which will play essential roles in our analysis below.

Our goal is to see the effects of very large super-horizon scale perturbations on smaller CMB scales perturbations. In this view, the size of our observed Universe is at the order of $H_0^{-1}$ in which $H_0$ is the current Hubble expansion rate while the  long mode (the modulating mode)   has the wavelength $\lambda_L \gg H_0^{-1}$. As usual, in order to perform the calculations  we go to the Fourier space with the volume $V$.  The volume of Fourier space $V$ should be larger than the whole universe. On the other hand, we also assume that the scale of long wavelength mode is much larger than the scale of Fourier space. As a result we have the following hierarchy
\ba
H_0^{-3} \ll V \ll k_L^{-3}
\ea
in which $k_L = 1/\lambda_L$ is the long mode comoving wave number.  As a result, the Fourier transform of long wavelength mode satisfies the following relation
\ba
\label{long-Fourier}
\calR_L(\x) \simeq \calR_{k_L} e^{i {\bf k_L.x}}
\ea
It is worth to mention that we have assumed one specific mode $k_L$ with a known wavelength in real space. Consequently there is a delta function term in Fourier  space integral $\delta(k-K_L)$, resulting in Eq. (\ref{long-Fourier}).  Now our aim is to parameterize the effects of $k_L$ on CMB scale modes $k$ such that
\ba
{\calR} = {{ \cal R}_{\bf k} }\left( 1 + A(k)  \hat {\bf p} . {\bf x}/x_n 
\right) \, ,
\ea
which is equivalent to Eq. (\ref{P-asym}). Note that curvature perturbation on logarithmic momentum scale $\calR_k$ is defined via
\ba
\label{R-power}
\langle \calR_{\bf k} \calR_{\bf{k'}} \rangle  = (2 \pi)^3\delta^3  ( {\bf{k} + \bf{k'}})  P_{\calR}(k)
\quad , \quad
{\cal P}_{\calR} \equiv \frac{k^3}{2 \pi^2}  P_{\calR}(k) \, .
\ea
Starting with the parameterizations Eq. (\ref{P-asym}) the fractional change due to long wavelength modulation on curvature perturbation power spectrum is
\ba
\label{grad}
\frac{\nabla P_{\calR_k}}{P_{\calR_k}} \simeq \frac{2 A \hat {\bf p}}{x_n}
\ea
which will be used later.

Now we can study how the local non-Gaussianity modifies the power spectrum of CMB-scale perturbations, leading to anisotropy. To start, let us assume that the curvature perturbations does not evolve on super-horizon scales, i.e. $\dot \calR_k =0$ for $k \ll a H$. This is true for all single field models in which the system has reached the attractor regime in which the decaying mode can be neglected. However, this assumption does not hold in models of non-attractor single field inflation \cite{Namjoo:2012aa, Chen:2013aj}
in which the would-be decaying mode actually dominates over the would-be growing mode, the constant mode. As a result, during the non-attractor regime $\dot \calR_k \neq 0$ even for single field inflation. We will study the general non-attractor case shortly.  Consider the three-point correlation function of the long wavelength mode with  two small (CMB scales) modes as follows
\ba
\label{rescale}
\Big \langle \calR({\bf k_L}) \calR( {\bf k_1}) \calR( {\bf k_2}) \Big \rangle
\simeq
 \Big{\langle} \, \, \calR({\bf k_L}) \,\,  \Big{\langle} \, \,   \calR( {\bf k_1})\calR({\bf k_2}) \Big\rangle_{\calR({\bf k_L})}  \Big{\rangle}
  \simeq
\Big \langle \calR({\bf k_L}) \,\, \left(   \calR({\bf k_L}) \dfrac{\partial}{\partial \calR({\bf k_L})} \Big \langle \calR({\bf k_1})\calR({\bf k_2}) \Big \rangle \large \Big\vert_{\calR({\bf k_L})=0}  \right)
 \Big\rangle
\ea
This  equation is the key for our analysis below. Intuitively this equation means  that the effects of a large scale perturbation would be just a rescaling of the background  for small scale perturbations. This is because the  small scale perturbations can not probe the spatial variation associated with the long wavelength mode. This logic was employed in \cite{Creminelli:2004yq, Ganc:2010ff, RenauxPetel:2010ty} to calculate the amplitude of the local non-Gaussianities and to check the Maldacena's consistency condition that $f_{NL} \sim (n_s -1) $.
Note that to write  Eq. (\ref{rescale}) we have assumed that there is only one degree of freedom encoded in $\calR$, so there is no other source of perturbation. Secondly, it is assumed that there is no intrinsic non-Gaussianity deep inside the horizon which means we have started from a Bunch-Davies vacuum. In other words, the non-Gaussianities generated in the model are either at the time of horizon crossing or  on the super-horizon scales. These two assumptions (the Bunch-Davies (BD) initial condition and the assumption of single source perturbations) 
are the only assumptions in our analysis in this section. However, note that we do not impose the slow-roll conditions. As long as a violation of slow-roll condition does not violate the  BD initial conditions for modes inside the horizon, then our consistency condition does hold. For example, this is the case when the violation of slow-roll conditions happens after the modes of interest has left the horizon. 

Note that Eq. (\ref{rescale}) may not be a practically useful method to calculate $f_{NL}$. We instead employ an inverse engineering method in which  we look for the amplitude of asymmetry, $A$, for a given value of $f_{NL}$. Therefore, to calculate the bispectrum one has to employ other methods, such as the in-in formalism or the $\delta N$
formalism.

Defining the bispectrum $B_\calR ({\bf k}_1, {\bf k}_2, {\bf k}_L)$ in the Fourier space via
\ba
\label{bi- def}
\langle \calR_{{\bf k}_1} \calR_{{\bf k}_2} \calR_ {{\bf k}_L} \rangle &\equiv& \left( 2 \pi \right)^3 \delta^3 \left({\bf k}_1 +  {\bf k}_2 +  {\bf k}_L \right) B_{\calR}({\bf k}_1 , {\bf k}_2, {\bf k}_L )
\ea
 Eq. \eqref{rescale} yields
\ba
\label{B-eq}
B_{\calR}({\bf k}_1 , {\bf k}_2, {\bf k}_L )  = P_{\calR_{k_L}} \dfrac{\partial P_{k_1}}{\partial \calR_L} \, .
\ea
On the other hand, the amplitude of the local non-Gaussianity  $f_{NL}$
in the squeezed limit $ k_L \ll k_1 \simeq k_2$ is given by
\ba
\label{fNL-def}
f_{NL}  = \lim_{k_L \rightarrow 0} \frac{5}{12}
\frac{B_{\calR}({\bf k}_1 , {\bf k}_2, {\bf k}_L )}{P_{\calR_{ {\bf k}_1 }}
P_{\calR_{ {\bf k}_L}}  } \, .
\ea
Comparing Eqs. (\ref{B-eq}) and (\ref{fNL-def}) yields
\ba
\dfrac{1}{P_{\calR_k}} \dfrac{\partial P_{\calR_k}}{\partial \calR_L} = \dfrac{12}{5} f_{NL}
\ea
in which it is understood that $k=k_1=k_2$ corresponds to the CMB-scale perturbations.

Since we are looking for the directional modification of power spectrum due to large scale perturbations we can write the above equation as
\ba
\label{nablaPzeta}
 \dfrac{1}{P_{\calR_k}} \nabla P_{\calR_k} = \dfrac{12}{5} f_{NL}  \nabla \calR_L \, .
\ea
Now comparing this equation with Eq. (\ref{grad}) yields the following formula for $A$
\ba
\label{A-eq}
A= \frac{6}{5} f_{NL} x_n  |\nabla \calR_L |
\ea
On the other hand,  using Eq. (\ref{long-Fourier}) for the decomposition of the long mode in the Fourier space, we have
 \ba
 \label{gradR}
 \vert \nabla \calR_L \vert \simeq  k_L   | \calR_L| =
 k_L {\cal P}^{1/2}_{\calR_L} = k_L E {\cal P}^{1/2}_{\calR} \, ,
 \ea
in which we have used $\calR_L = {\cal P}^{1/2}_{\calR_L}$, and following the Lyth convention \cite{Lyth:2013vha},  we have introduced the parameter $E$ via
\ba
\label{E-def}
{\cal P}_{\calR_L} = E^2 {\cal P}_\calR \, .
\ea
Here it is understood that $\calR$ is the CMB-scale curvature perturbation with the power spectrum ${\cal P}_\calR$. Now plugging the value of $\vert \nabla \calR_L \vert$ obtained 
in Eq. (\ref{gradR}) into Eq. (\ref{A-eq}) yield our desired formula 
\ba
\label{AfNL}
A(k) = \dfrac{6}{5} f_{NL} \, k_L x_n E \, {\cal P}^{1/2}_{\calR_k}
\ea
This is our consistency condition, relating the amplitude of bipolar hemispherical asymmetry
to the amplitude of non-Gaussianity in the squeezed limit. Note that this formula applies for all single source models of inflation with the BD initial conditions  which have reached the attractor limit, i.e. with the assumption $\dot \calR =0$ on the super-horizon scales. Note that Eq. (\ref{AfNL}) was  first obtained by Lyth \cite{Lyth:2013vha} for the curvaton scenario. 

Having obtained our consistency relation for the attractor models, now we proceed with the analysis for the general single source inflationary models including the non-attractor models as
studied in \cite{Namjoo:2012aa, Chen:2013aj}. The analysis go parallel to the above analysis, except that in our starting point, Eq. (\ref{rescale}), we have to also consider the modulation of the two-point functions of small scales by $\dot \calR$. Assuming that both $\calR$ and
$\dot{\calR_L}$ are independent variables, and  following the same procedures as before, yields
\ba
\label{rescale2}
\Big \langle \calR({\bf k_L}) \calR( {\bf k_1}) \calR( {\bf k_2}) \Big \rangle
\simeq
\Big \langle \calR({\bf k_L}) \,\, \left(   \calR_L({\bf k_L}) \dfrac{\partial}{\partial \calR({\bf k_L})} \Big \langle \calR({\bf k_1})\calR({\bf k_2}  ) \Big \rangle \large 
+  \dot\calR({\bf k_L}) \dfrac{\partial}{\partial \dot\calR({\bf k_L})} \Big \langle \calR({\bf k_1})\calR({\bf k_2}) \Big \rangle \large   \,  \right)
 \Big\rangle
\ea
As before,  using the definition of the bispectrum from Eq. (\ref{bi- def}), this equation can be further manipulated to give
\ba
\label{squeeze-dotzeta}
\dfrac{12}{5} f_{NL} P_{\calR_L} P_{\calR} \simeq
\Big\langle  \calR_L \left( \calR_L \dfrac{\partial P_\calR}{\partial \calR_L}
+\dot \calR_L \dfrac{\partial P_\calR}{\partial \dot \calR_L} \right)   \Big\rangle
=
P_{\calR_L} \dfrac{\partial P_\calR}{\partial \calR_L}  + \dfrac{1}{2} \partial_t{P_{\calR_L}} \dfrac{\partial P_\calR}{\partial \dot \calR_L}
\ea
Now we need
\ba
\label{eq1}
\nabla P_\calR =
 \dfrac{\partial P_\calR}{\partial \calR_L} \nabla \calR_L + \dfrac{\partial P_\calR}{\partial \dot \calR_L}
 \nabla \dot{\calR_L}
\ea
Noting that $\nabla \calR_L = k_L \calR_L$ and $\calR_L = {\cal P}_{\calR_L}^{1/2}$ we conclude
\ba
\label{eq2}
\calR_L \nabla P_\calR =
 \dfrac{\partial P_\calR}{\partial \calR_L} k_L {\cal P}_{\calR_L} + \dfrac{1}{2}\dfrac{\partial P_\calR}{\partial \dot \calR_L}
 k_L  \dot{\cal P}_{\calR_L} \, .
\ea
Comparing this with \eqref{squeeze-dotzeta} yields
\ba
\label{eq3}
\dfrac{\nabla P_\calR}{P_\calR} \simeq \dfrac{12}{5} f_{NL} k_L  {\cal P}_{\calR_L}^{1/2}
= \dfrac{12}{5} f_{NL} \nabla \calR \, .
\ea
Interestingly, this is exactly the same as  \eqref{nablaPzeta} and following the same steps as before, we obtain our consistency relation Eq. (\ref{AfNL}). Note that in above calculations,
in going from Eq. (\ref{eq1}) to Eq. (\ref{eq3}), we have used the fact that $\calR_L$ and $\dot \calR_L$ are in phase, i.e. $\calR_L({\bf x}) = \calR_{k_L}(t) e^{-i k_L.{\bf x}}$ and $\dot \calR_L({\bf x}) = \dot \calR_{k_L}(t) e^{-i k_L. {\bf x}}$ which follows from our long wavelength mode decomposition
given in Eq. (\ref{long-Fourier}).

In summary, we have the consistency relation given by Eq. (\ref{AfNL}) valid for all models of inflation in which the perturbations are produced by a single source with the BD initial condition.  This result is independent of the slow-roll assumptions and also is valid for non-attractor models alike.
Furthermore,   this result is independent of the number of background fields which can contribute to the background inflationary expansion but do not contribute to the curvature perturbations. This include the standard curvaton scenario in which the inflaton does not contribute to the curvature perturbations.  

To obtain  useful information from the abstract formula Eq. (\ref{AfNL}) we need to have some information on the magnitude of the combination $ k_L x_n E {\cal P}_{\calR}^{1/2}$.  The Observational constraints from the quadrupole and octupole moments of anisotropic power spectrum of CMB can be used  to constrain this combination. To do this, note that  we are considering a specific mode of long wavelength modulation (i.e. ${\cal{R}_L}=E{\cal{P}}^{1/2}_{\cal{R}}\sin(k_L.x+\omega)$) and its gradient in the CMB is 
$\nabla {\cal{R}}_L\simeq |(kx_n)E{\cal{P}}^{1/2}_{\cal{R}}\cos(\omega)|$ with $\omega$ being an arbitrary phase.  In Erickcek et al. \cite{Erickcek:2008jp} it was shown that the quadrupole and octupole constraints are as below:
 \begin{eqnarray}\label{qpole}
 (kx_n)^2E{\cal{P}}^{1/2}_{\cal{R}}\sin\omega <  (5/3)\times 5.8Q_2\sim  (5/3) \times 5.8 \times (6\times 10^{-6}) \\ \nonumber
 (kx_n)^3E{\cal{P}}^{1/2}_{\cal{R}}\cos\omega <   (5/3) \times  32Q_3\sim (5/3)\times 32 \times (9\times 10^{-6}) 
 \end{eqnarray}
 where the additional factor $5/3$ comes from exchanging the gravitational potential $\Phi$ to curvature perturbation $\calR$. We note that the phase appearing in quadrupole and octupole are the same as in the long wavelength mode. The octupole can constrain the  change of the long wavelength mode independent of the phase, while the quadrupole  vanishes for $\omega=0$. Consequently, the constraints on the power of long wavelength comes from octupole term.   On the other,  hand the amplitude of long mode perturbation inside the big box, i.e. the super-universe, is  $ \calR_L=  E {\cal P}_{\calR}^{1/2}$. In order for this to be consistent  as a perturbations, we  require 
\ba
\label{bound2}
 \calR_L^2  \simeq  E^2 {\cal P}_{\calR}(k_1) \lesssim 1.
\ea
Combining the bounds from Eqs. (\ref{qpole}) and (\ref{bound2}) yields
\ba
\label{bound3}
\frac{6}{5}(k_L x_n) \,  E {\cal P}_{\calR}^{1/2} \lesssim  {32Q_3}^{1/3} \sim 10^{-1}
\ea
Plugging this upper bound into Eq. (\ref{AfNL}) , we obtain the following {{\it{upper bound consistency condition}}  for the amplitude of dipolar asymmetry 
\ba
\label{A-upper}
| A |  \lesssim 10^{-1} | f_{NL} | \, .
\ea
To explain the observed anisotropy we require $\vert A(k) \vert =0.07 \pm 0.02$.
Now we employ our result to some interesting examples. 

First consider the standard slow-roll single field inflation. From Maldacena's analysis \cite{Maldacena:2002vr} (see also \cite{Kundu:2011sg}) 
it is known that $f_{NL} \sim (n_s -1)$. Using this in Eq. (\ref{A-upper}) gives $A \lesssim 10^{-1} (n_s - 1) \sim 10^{-3}$, 
as observed in \cite{Erickcek:2008sm,  Lyth:2013vha}( see also \cite{Liu:2013kea} in different context). This is too small to explain the observed
dipolar asymmetry.
Now consider the standard curvaton scenario, in which all perturbations are sourced by the curvaton field. In this model, $f_{NL}$ is uncorrelated to $n_s-1$ so one can easily saturate the observational bound from Planck $f_{NL} = 2.7\pm 5.8$ (68 \% CL) \cite{Ade:2013ydc}.
As a result one can easily saturate the observational bound  $\vert A(k) \vert =0.07 \pm 0.02$ on the CMB scale by taking $\vert f_{NL} \vert \lesssim 5$  \cite{Lyth:2013vha}. However,  $f_{NL}$ is nearly scale-invariant for the standard curvaton scenario so this model may not work easily in order to reduce the amplitude of asymmetry on small scales as required from the quasar constraints.

As a non-trivial example, we consider the non-attractor single field models studied in \cite{Namjoo:2012aa, Chen:2013aj}. In these models, inflation has two stages: the non-attractor phase at the early stage of inflation followed by the attractor regime towards the end of inflation. During the non-attractor phase the curvature perturbation is not frozen so $\dot \calR$ is not zero on super-horizon scales. As observed in \cite{Namjoo:2012aa, Chen:2013aj} a large value of local non-Gaussianity can be generated given by
\ba
\label{f-NLcs}
f_{NL} =\frac{5(1+ c_s^2)}{4 c_s^2} \, ,
\ea
in which $c_s$ is the sound speed of perturbations during the non-attractor phase. Choosing
small enough $c_s$ one can easily saturate the observational bound on $f_{NL}$. For example  with $ f_{NL} \lesssim 2$ one can easily find large observable bipolar asymmetry. One interesting aspect of this model is that  for the modes which leave the horizon during the attractor phase the curvature perturbation is frozen so $\dot\calR\simeq 0$ once these modes leave the horizon.
This means that $f_{NL}$ has a built-in scale-dependence in non-attractor scenarios. For modes which leave the horizon during the non-attractor phase $f_{NL}$ is given by Eq. (\ref{f-NLcs}) while for the modes leaving the horizon during the attractor phase $f_{NL} \simeq n_s -1 \sim 0$. This helps to satisfy both the CMB constraints and the quasar constraints on $A(k)$. For this picture to work, we tune the parameters such that the observed CMB scales leave the horizon during the first 5-10 numbers of e-folds when the system is in non-attractor phase. Scales smaller than $Mpc^{-1}$ are supposed to leave the horizon once the background reaches the attractor regime during last 50 or so e-folds. By construction $f_{NL}$ and consequently  $A(k)$ are reduced sharply for these scales. It will be an  interesting exercise to fit the predictions of this scenario with the Planck data.
\subsection{$\delta N$ formalism for single source models}

As a further illustration, here we employ  the $\delta N$ formalism \cite{Sasaki:1995aw, Wands:2000dp} to obtain Eq. (\ref{AfNL}) for a generic single source inflationary model.
Suppose the perturbations are generated by the single source (in the sense defined above)
$\delta \phi$.  We have to Extend the usual $\delta N$ formalism to more general case in which the background number of e-fold $N$ is defined in the phase space \cite{Namjoo:2012aa, Chen:2013aj} so $N= N(\phi, \dot \phi)$. Note that the dependence of $N$ to $\dot \phi$ is crucial in non-attractor models as emphasized in \cite{Namjoo:2012aa}, otherwise one obtains a wrong result for $\calR$.  Expanding $N$ to second order in perturbations of $\delta \phi$ and $\delta \dot \phi$,  the comoving curvature perturbation to second order is given by
\ba
\label{R-alpha}
\calR = N_{,\phi} \delta \phi +  N_{,\dot \phi}  \delta \dot \phi +
\dfrac{1}{2} N_{,\phi \phi} \delta \phi^2 +
\dfrac{1}{2}  N_{,\dot \phi \dot \phi}  {\delta \dot \phi}^2 +
 N_{,\phi \dot \phi} \delta \phi \delta \dot \phi \, .
\ea
In this view both $\delta \phi$ and $\delta \dot \phi$ are treated independent random variables.
However, in separate Universe approach $\delta \phi$ and $\delta \dot \phi$ are classical fields
measuring the deviation from a given background solution. In the gradient expansion limit where both $\delta \phi$ and $\delta \dot \phi$ becomes classical, one expect that $\delta \dot \phi$ carries the same initial quantum (statistical) information as $\delta \phi$. As a result, to leading order in gradient expansion $\delta \phi$ and $\delta \dot \phi$ are related to each other via $\delta \dot \phi = \alpha(t) \delta \phi$ in which $\alpha(t)$ is a function depending only on background quantities. For example in models of slow-roll inflation in which $\delta \phi$ freezes out on super-horizon scales $\alpha=0$. On the other hand, for the non-attractor model studied in \cite{Chen:2013aj}
$\delta \phi = \delta \phi_0 e^{\alpha N}$ with $\alpha$ a constant.  With this assumption in mind, from Eq. (\ref{R-alpha}) one obtains
\ba
\dfrac{6}{5} f_{NL} = \dfrac{N_{,\phi \phi} + 2\alpha N_{,\phi \dot \phi} +\alpha^2 N_{,\dot \phi \dot \phi} }{\left(N_{,\phi} + \alpha N_{,\dot \phi} \right)^2} \, .
\ea
Note that to obtain this formula, we do not to know the specific form of $\alpha (t)$. All we need is that  $\alpha (t)$ does not carry independent  quantum (statistical) information.  

On the other hand, expanding the coefficients of  linear parts of $\calR$ in Eq. (\ref{R-alpha})
as a function of background field modulated by the long mode $\delta \phi_L$, i.e.
$N_{,\phi} (\phi+ \delta \phi_L)= N_{,\phi} (\phi) + \delta \phi_L N_{, \phi \phi} $, one obtains
\ba
\label{R-deltaN}
\calR &\simeq& \calR^{iso} \left( 1+
\dfrac{N_{,\phi \phi} + 2\alpha N_{,\phi \dot \phi} +\alpha^2 N_{,\dot \phi \dot \phi}}{N_{,\phi} + \alpha N_{,\dot \phi}}    \,  \delta \phi_L  \right) \nonumber\\
&=&  \calR^{iso} \left( 1+ \frac{6}{5} f_{NL} \left(N_{,\phi} + \alpha N_{,\dot \phi} \right)
\delta \phi_L \right) \, .
\ea
Following the decomposition of long wavelength mode in Eq. (\ref{long-Fourier}) we have
\ba
\delta \phi_L  ({\bf x})= {\cal P}^{1/2}_{\delta \phi_L} e^{i {\bf k}_L. {\bf x}} \, .
\ea
Plugging this into Eq. (\ref{R-deltaN}) yields
\ba
{\cal P}_{\cal R} = {\cal P}_{\cal R}^{iso} \left\vert 1+ \frac{6}{5} f_{NL} \left(N_{,\phi} + \alpha N_{,\dot \phi} \right) \delta  {\cal P}^{1/2}_{\delta \phi_L} e^{i {\bf k}_L. {\bf x}}
\right\vert^2
\ea
As a result, the fractional change in the gradient  is
\ba
\label{grad2}
\frac{\nabla {\cal P}_{\cal R}}{{\cal P}_{\cal R}}= \frac{12}{5} f_{NL} k_L
 \left(N_{,\phi} + \alpha N_{,\dot \phi} \right) {\cal P}^{1/2}_{\delta \phi_L} \, .
\ea
As before, we have to relate the power spectrum of the long mode, ${\cal P}^{1/2}_{\delta \phi_L}$, to the power spectrum of the short mode, ${\cal P}^{1/2}_{\delta \phi}$, via
${\cal P}^{1/2}_{\delta \phi_L} = E {\cal P}^{1/2}_{\delta \phi}$. Plugging this relation in Eq. (\ref{grad2}), using the identity ${\cal P}^{1/2}_{\delta \phi} = {\cal P}^{1/2}_{\calR}/(N_{,\phi} + \alpha N_{,\dot \phi}) $  and comparing with the definition given in Eq. (\ref{grad}) yields
\ba
A(k) = \dfrac{6}{5} f_{NL} \, k_L x_n E \, {\cal P}^{1/2}_{\calR_k} \, .
\ea
Interestingly, this is exactly the same as Eq. (\ref{AfNL}).


\section{Asymmetry from Multiple Fields Models}
\label{multiple}

In this section we study the asymmetry generated in models of multiple fields inflation in which more than one field contributes in curvature perturbations. As mentioned before, the example of curvaton was studied in the literature and it is argued that the asymmetry generated from curvaton scenarios with a mixture of inflaton and curvaton contributions to the curvature perturbations can generate observable bipolar asymmetry \cite{Erickcek:2008jp, Erickcek:2009at, Lyth:2013vha}.  Here we consider the general case
of multiple fields scenarios in which the quantum fluctuations of more than one light scalar fields source the power spectrum and the bispectrum. The goal is to see if the super-horizon modulations of these light scalar fields can generate hemispherical asymmetry and to find the relation between the amplitude of bispectrum and the amplitude of the asymmetry. For a related
recent work see \cite{McDonald:2013aca}. 

To be specific, we consider the model containing two light scalar fields $\phi$ and $\sigma$ which contribute into both background and the perturbations dynamics. The generalization to
more than two fields is straightforward.  Employing the standard  $\delta N$ formalism for super-horizon perturbations, the curvature perturbation is given by
\ba
\calR = N_{, \phi} \delta \phi + N_{, \sigma} \delta \sigma + \dfrac{1}{2} N_{, \phi \phi} \delta \phi^2 + \dfrac{1}{2} N_{, \sigma \sigma} \delta \sigma^2 + N_{, \phi \sigma} \delta \phi \delta \sigma
\ea
To simplify the analysis, we assume that $N$ does not depend to $\dot \phi$
and $\dot \sigma$ so the system has reached the attractor regime. 
The power spectrum is given by
\ba
\calP_\calR = N_{, \phi}^2 \calP_ {\delta \phi} + N_{, \sigma}^2 \calP_{\delta \sigma}  \, ,
\ea
in which $\calP_ {\delta \phi}$ and $\calP_ {\delta \sigma}$ represents the power spectrum of light fields quantum fluctuations $\calP_ {\delta \phi} = \calP_ {\delta \sigma} =(H/2\pi )^2$. However, one may consider the general case in which the light scalar fields do not have equal power spectrum.

Following the notation of \cite{Fonseca:2012cj, Assadullahi:2013ey} (see also \cite{Langlois:2013dh}), the non-Gaussianity can be written as the weighted sum of each field's contribution
\ba
\label{fNL-two}
f_{NL} = w_\phi^2 f_{NL}^{\phi} +w_\sigma^2 f_{NL}^{\sigma} + 2 w_\phi w_\sigma f_{NL}^{\phi \sigma}
\ea
where, we have defined
\ba
\dfrac{6}{5} f_{NL}^{\phi} \equiv \dfrac{N_{, \phi \phi}}{N_{,\phi}^2 }
\qquad , \qquad
\dfrac{6}{5} f_{NL}^{\sigma} \equiv \dfrac{N_{, \sigma \sigma}}{N_{,\sigma}^2 }
\qquad , \qquad
\dfrac{6}{5} f_{NL}^{\phi \sigma} \equiv \dfrac{N_{, \phi \sigma}}{N_{,\sigma} N_{,\phi } }
\ea
and $w_\sigma$ and $w_\phi$ are the fractional contribution of fields $\sigma$ and $\phi$ to the power spectrum, respectively.
\ba
\label{w}
w_\phi = \dfrac{N_{,\phi}^2 \calP_{\delta \phi}}{\calP_\calR}= \frac{N_{,\phi}^2}{N_{,\phi}^2+ N_{,\sigma}^2}
 \qquad , \qquad w_\sigma = 1-w_\phi \, .
\ea
where the last equality in $w_\phi$ holds if we assume $\calP_ {\delta \phi} = \calP_ {\delta \sigma} =(H/2\pi )^2$.

Now we can obtain the anisotropy of the model. Assuming a modulation of the background by two large scale modes $\delta \sigma_L$ and $\delta \phi_L$, one has
\ba
\label{aniso-expand}
\calR = N_{,\phi} \delta \phi + N_{,\sigma} \delta \sigma +  N_{, \phi \phi} \delta \phi \delta \phi_L +  N_{, \sigma \sigma} \delta \sigma \delta \sigma_L
+ N_{, \phi \sigma} (\delta \phi \delta \sigma_L +\delta \phi_L \delta \sigma )
\ea
For large scale perturbations we have
\ba
\delta \phi_L(x) = \calP_{\delta \phi_L }^{1/2} e^{-i {\bf k_L.x}}
\qquad , \qquad
\delta \sigma_L(x) = \calP_{\delta \sigma_L }^{1/2} e^{-i {\bf k_L.x}}
\ea
As before, we can relate the power spectrum of the large scale to the power spectrum of small scale via
\ba
\calP_{\delta \phi_L} = E^2 \calP_{\delta \phi}
\qquad , \qquad
\calP_{\delta \sigma_L} = E^2 \calP_{\delta \sigma}  \, ,
\ea
where we have assumed that the enhancement factors are the same for both fields. This seems a reasonable assumption  because the dynamics of generating the light scalar fields quantum fluctuations deep inside the horizon from an initial Bunch-Davies vacuum and the subsequent horizon crossing effects  are the same for both light fields. However, one may consider the general case in which the enhancement factor $E$ amy be different for the two scalar fields. 
Therefore, we have
\ba
\calR =  \left[ N_{,\phi} + E \left( N_{, \phi \phi}  \calP_{\delta \phi}^{1/2}
+ N_{, \phi \sigma}  \calP_{\delta \sigma}^{1/2}\right)  e^{-i {\bf k_L.x}} \right] \delta \phi
+
 \left[ N_{,\sigma} + E \left( N_{, \sigma \sigma}  \calP_{\delta \sigma}^{1/2}
+ N_{, \phi \sigma}  \calP_{\delta \phi}^{1/2}\right)  e^{-i {\bf k_L.x}} \right] \delta \sigma
\ea
Using this form of $\calR$, one can check that the directional dependence of $\calP_{\calR}$ is given by
\ba
\Big\vert \dfrac{\nabla \calP_\calR}{\calP_\calR} \Big\vert= \dfrac{12}{5} E  \calP_\calR^{1/2}
\left[ f_{NL}^{\sigma \phi} (w_\sigma w_\phi^{1/2}\, sgn(\dfrac{N_{,\phi}}{N_{,\sigma}})+ w_\sigma^{1/2} w_\phi ) +
f_{NL}^{\phi} w_\phi^{3/2} \, sgn(\dfrac{N_{,\phi}}{N_{,\sigma}})
+f_{NL}^{\sigma} w_\sigma^{3/2}
   \right] \, .
\ea
where $sgn$ is the sign funtion, $+1$ for postive argument and $-1$ for negative. Comparing this with the anisotropic ansatz for power spectrum asymmetry Eq. (\ref{grad}) yields
\ba
\label{A-two}
A &=&   \dfrac{6}{5} x_n k_L E  \calP_\calR^{1/2}
\left[ f_{NL}^{\sigma \phi} (w_\sigma w_\phi^{1/2} \, sgn(\dfrac{N_{,\phi}}{N_{,\sigma}})+ w_\sigma^{1/2} w_\phi ) +
f_{NL}^{\phi} w_\phi^{3/2} \, sgn(\dfrac{N_{,\phi}}{N_{,\sigma}})
+f_{NL}^{\sigma} w_\sigma^{3/2}  \right] \, \nonumber\\
&\lesssim& 10^{-1} \left[ f_{NL}^{\sigma \phi} (w_\sigma w_\phi^{1/2} \, sgn(\dfrac{N_{,\phi}}{N_{,\sigma}})+ w_\sigma^{1/2} w_\phi ) +  f_{NL}^{\phi} w_\phi^{3/2}\, sgn(\dfrac{N_{,\phi}}{N_{,\sigma}}) +f_{NL}^{\sigma} w_\sigma^{3/2}  \right] \, .
\ea
in which we have used the octupole constraint Eq. (\ref{bound3}) to eliminate the factor 
$\dfrac{6}{5} x_n k_L E  \calP_\calR^{1/2}$.
This is a very interesting result which is also intuitively reasonable.  The amplitude of asymmetry depends on how much each field contributes to the curvature perturbation power spectrum, parameterized by $w^\phi$ and $w^\sigma$,  and also on how strong the coupling between the small modes to the large modes for each field's perturbations is, as parametrized by the  corresponding $f_{NL}$. Interestingly, we also see the effects of coupling of small modes to the large modes with two different sources, as indicated by the term $f_{NL}^{\sigma \phi}$
in Eq. (\ref{A-two}). Note that our result Eq. (\ref{A-two})  is different than the result obtained by Erickcek et al. \cite{Erickcek:2008sm} in the particular case of curvaton in which they found $A \propto f_{NL}^{1/2}$.

As discussed in \cite{Wang:2013lda} one can chose the signs of $f_{NL}^{\phi}, f_{NL}^{\sigma}, f_{NL}^{\phi \sigma}$ and the weights of each field in power spectrum
such that the  total $f_{NL}$ in Eq. (\ref{fNL-two}) can be small enough to satisfy the Planck observations $f_{NL} = 2.7\pm 5.8$ (68 \% CL) \cite{Ade:2013ydc}.
Then one still has  enough free parameters in Eq. (\ref{A-two}) to make $A$ large enough to saturate the observational bounds deduced from Planck. On top of this, one should also check the observational constraints on the amplitude of tri-spectrum $\tau_{NL}$ which in this model
is calculated to be \cite{Assadullahi:2013ey}
\ba
 \label{tauNL-two}
\frac{25}{36} \tau_{NL} = w_\sigma^3 (f_{NL}^\sigma)^2 +2w_\sigma^2 w_\phi f_{NL}^\sigma f_{NL}^{\sigma\phi}  
 + w_\sigma w_\phi ( w_\sigma + w_\phi ) (f_{NL}^{\sigma\phi} )^2 +
2 w_\sigma w_\phi^2 f_{NL}^{\sigma\phi} f_{NL}^\phi + w_\phi^3 (f_{NL}^\phi)^2 \,
\ea
The upper bound inferred from Planck data is \cite{Ade:2013ydc} $\tau_{NL} < 2800$ (95 \% CL). It would be an interesting exercise to search for parameter values in which
large observable value of $A$ can be obtained from   Eq. (\ref{A-two}) while  the  Planck constraints on $f_{NL}$ and $\tau_{NL}$, given respectively by
Eq. (\ref{fNL-two}) and Eq. (\ref{tauNL-two}), are satisfied.

As an interesting example, consider the case in which the inflaton field $\phi$ is Gaussian so
$f_{NL}^\phi \simeq f_{NL}^{\phi\sigma}\simeq0$. From Eq. (\ref{fNL-two}) one obtains
$f_{NL}^\sigma \simeq f_{NL}/w_\sigma^2$ so Eq. (\ref{A-two}) yields 
\ba
\label{A-fNL0}
A \lesssim \frac{f_{NL}}{10 \sqrt{w_\sigma}} \, .
\ea
This is a very interesting result. If one reduces the contribution of the non-Gaussian field $\sigma$ in the power spectrum, i.e. $w_\sigma \ll 1$, with $f_{NL} $ at the order of unity 
one can get large upper bound on the amplitude of asymmetry.  

It  may be more transparent to eliminate $w_\sigma$ in terms of $\tau_{NL}$ in  Eq. (\ref{A-fNL0}). Using Eq.  (\ref{tauNL-two}), one obatins
\ba
\label{A-tauNL}
A \lesssim \frac{\sqrt{\tau_{NL}}}{12} \, .
\ea
It is very interesting that the upper bound on $A$ is independently controlled by $\tau_{NL}$.
With the upper bound  $\tau_{NL} < 2800$ (95 \% CL) from Planck data  \cite{Ade:2013ydc}  one obtains $A \lesssim 4.4$.

\section{Asymmetry from inhomogeneities generated at the End of Inflation}

As a concrete example, in this section we study bipolar asymmetry generated from inhomogeneities generated at the surface of end of inflation. This model was studied in \cite{Bernardeau:2004zz, Dvali:2003em, Lyth:2005qk} as a mechanism to provide large local non-Gaussianity. Here we follow the model studied in \cite{Lyth:2005qk} and \cite{Assadullahi:2012yi}.

The model is based on inhomogeneities generated from the modulation of the surface of end of inflation in the background of a hybrid inflation \cite{Linde:1993cn, Copeland:1994vg} scenario. The model contains three fields, the inflaton field $\phi$, the waterfall field $\chi$ and the light isocurvature field $\sigma$. The potential is
\ba
\label{potential}
V= \frac{\lambda}{4} \left( \chi^2 - \frac{M^2}{\lambda} \right)^2  + \frac{m^2}{2} \phi^2
+ \frac{g^2}{2} \phi^2 \chi^2 + \frac{\gamma^2}{2} \chi^2 \sigma^2 \, .
\ea
In this picture, inflation proceeds as in usual hybrid inflation models. The waterfall field $\chi$
is very heavy so it quickly rolls down to its local minimum $\chi=0$, so during the whole period of inflation we set $\chi=0$. Inflation ends when the inflaton field reaches the surface of end of inflation defined by
\ba
 \label{end-surface}
 \phi_e^2 + \frac{\gamma^2}{g^2} \sigma^2 = \phi_c^2 \, ,
\ea
in which $\phi_e$ indicates the value of $\phi$ at the surface of end of inflation and the critical value $\phi_c$ is  defined by
\ba
\phi_c \equiv \frac{M}{g} \,.
\ea
Once $\phi$ hits the surface of end of inflation, the waterfall fields becomes tachyonic terminating inflation very quickly. As demonstrated in recent literature \cite{Lyth:2010ch, Abolhasani:2010kr, Fonseca:2010nk, Abolhasani:2011yp, Gong:2010zf, Clesse:2010iz, Martin:2011ib, Mulryne:2011ni} the waterfall quantum fluctuations do not contribute to large scale curvature perturbations.

The idea in \cite{Bernardeau:2004zz, Dvali:2003em, Lyth:2005qk} is to generated inhomogeneities from $\delta \sigma$ quantum fluctuations at the surface of end of inflation which also translate into  $\delta \phi$ perturbations in Eq. (\ref{end-surface}). With these descriptions, the final result for the curvature perturbation at the end of inflation, $\zeta_e$,
is the following (see \cite{Assadullahi:2012yi} for details).
\ba
 \label{final-zeta}
\zeta_e
 &=& \dfrac{3\delta\phi_\ast}{\alpha\phi_\ast} + \dfrac{3}{\alpha } \dfrac{\gamma^2 \sigma}{g^2 \phi_e^2}\delta\sigma +
  \dfrac{3}{2 \alpha} \dfrac{\gamma^2}{g^2 \phi_e^2}
  \left( 1+2 \dfrac{\gamma^2 \sigma^2}{g^2 \phi_e^2}  \right) \delta\sigma^2 +\ldots
  \nonumber \,,\\
 &=& \dfrac{3}{\alpha} \left[ \frac{\delta\phi_\ast}{\phi_\ast} + \frac{\calF}{1-\calF} \frac{\delta\sigma}{\sigma}
  + \frac{{\calF} (1+\calF)}{2(1-\calF)^2} \left( \frac{\delta\sigma}{\sigma} \right)^2 \right] +\ldots \,.
\ea
in which  we have defined
\ba
\label{alpha-beta}
\alpha \equiv \dfrac{m^2}{H^2}=\dfrac{12 \lambda m^2 M_P^2}{ M^4}
\ea
and
\be
 \label{defF}
 {\cal F} \equiv \dfrac{\gamma^2 \sigma_*^2}{g^2 \phi_c^2} \, .
\ee
Note that in Eq. (\ref{final-zeta}) we have neglected the second order term from $\delta \phi^2$
since the non-Gaussianities generated by inflaton is very small. Also note that $\alpha$ measures the mass of the inflaton compared to $H$ and as usual we assume that $\alpha \ll 1$ in order to get long enough period of inflation. Furthermore, the parameter $\calF$ defined above measures the contribution of $\sigma$ in energy density at the end of inflation, see \cite{Assadullahi:2012yi} for details, so
$0 \leq \calF <1$.

Assuming that the two fields $\delta \phi$ and $\delta \sigma$ are canonically normalized with the same amplitude, the contribution of the modulator to the power spectrum defined by Eq.\eqref{w} is here given by
\ba
w_\sigma = \dfrac{\calF^2 \phi_*^2}{\calF^2 \phi_*^2+(1-\calF)^2}
\ea
and the resulting non-Gaussianity, using Eq.\eqref{fNL-two} is the following
\ba
\dfrac{6}{5} f_{NL} = \dfrac{6}{5} w_\sigma^2 f_{NL}^\sigma
\ea
with
\ba
\label{fNL-lyth}
\dfrac{6}{5} f_{NL}^\sigma =  \, \dfrac{\alpha \, (1+\calF)}{ 6 \calF}
\ea
For fixed $w_\sigma$, and $\calF \ll 1$ we obtain a large non-Gaussianity.

Plugging the above value of $f_{NL}$ in  Eq. (\ref{A-two}),  the  upper bound on the amplitude of asymmetry is
\ba
\vert A \vert \lesssim 10^{-1} \times w_\sigma^{3/2} \,\dfrac{5 \alpha \, (1+\calF)}{ 36 \calF}
\ea
So we can obtain large asymmetry by requiring $\calF \ll 1$. Note that, using \eqref{fNL-lyth}, the last equation can be rewritten by
\ba
\label{A-sigma}
 \vert A \vert \lesssim   \dfrac{\vert f_{NL} \vert}{10 \, \sqrt{w_\sigma}} \, .
\ea
This is identical to Eq. (\ref{A-fNL0}) which is expected since in this model the inflaton field is Gaussian and the non-Guassianity is generated from the light field $\sigma $ by modulating the surface of the end of inflation. Similarly, eliminating $w_\sigma$ in Eq. (\ref{A-sigma})
in favor of $\tau_{NL}$ yields Eq. (\ref{A-tauNL}). As described below Eq. (\ref{A-tauNL}), large upper bound  on $A$ is obtained with the current Planck constraints on $\tau_{NL}$.\\


To summarize, the observed large scale anomalies in CMB may  indicate to a pre-inflationary physics. In this work we have shown that a long wavelength super-horizon mode  with a large amplitude can generate anisotropies on CMB scales, which can be related to the local type non-Gaussianity. In this work we have found an upper bound consistency relation between the anisotropy amplitude seen by Planck and the non-Gaussianity obtained from single source inflationary models. We have used the bounds from octupole moment of CMB as a crosscheck to find the upper bound consistency relation. We have shown that a large upper bound on $A$
is obtained in  single field non-attractor models of inflation which is consistent with the observational constraints from CMB and LSS (i.e. quasars).
This is because in these models the usual consistency condition $f_{NL} \sim 1-n_s$ does not hold and large observable non-Gaussianity can be generated.

We also studied the asymmetry generated in  models of multiple fields inflation in which more than one light field contribute to the curvature perturbation. It is shown that the amplitude of hemispherical asymmetry is controlled by the weighted sum of 
non-Gaussianity generated by each field. In the special case in which only one field contributes to non-Gaussianity while the other field is Gaussian, the amplitude of bipolar asymmetry is controlled by $\tau_{NL}$ as shown in Eq. (\ref{A-tauNL}). In particular, if one reduces the contribution of the non-Gaussian field in the curvature perturbation a large upper bound on
$A$ is obtained as shown in   Eq. (\ref{A-fNL0}). As a concrete example of models of this type  we studied asymmetry generated from inhomogeneities from the modulation of the surface of end of inflation. We  have found  that one can generate an observable asymmetry with $f_{NL}\sim {\cal{O}}(1)$ if the contribution of the modulator field to the power spectrum,  $w_\sigma$, is sufficiently small.

 \section*{Acknowledgment}
We would like to thank Adrienne L. Erickcek, Eiichiro Komatsu, David Lyth,  Misao Sasaki, David Wands and Lingfei Wang for many insightful comments and discussions.


{}


\section*{References}



\begin{thebibliography}{}		


\bibitem{Ade:2013zuv}
  P.~A.~R.~Ade {\it et al.}  [Planck Collaboration],
  ``Planck 2013 results. XVI. Cosmological parameters,''
  arXiv:1303.5076 [astro-ph.CO].


\bibitem{Guth:1980zm}
  A.~H.~Guth,
  ``The Inflationary Universe: A Possible Solution to the Horizon and Flatness Problems,''
  Phys.\ Rev.\ D {\bf 23}, 347 (1981).
;
  A.~A.~Starobinsky,
  Phys.\ Lett.\ B {\bf 91}, 99 (1980).
; K.~Sato,
  ``First Order Phase Transition of a Vacuum and Expansion of the Universe,''
  Mon.\ Not.\ Roy.\ Astron.\ Soc.\  {\bf 195}, 467 (1981).
  Phys.\ Lett.\ B {\bf 108}, 389 (1982).
;
  A.~Albrecht and P.~J.~Steinhardt,
  ``Cosmology for Grand Unified Theories with Radiatively Induced Symmetry Breaking,''
  Phys.\ Rev.\ Lett.\  {\bf 48}, 1220 (1982).


\bibitem{Ade:2013uln}
  P.~A.~R.~Ade {\it et al.}  [Planck Collaboration],
  ``Planck 2013 results. XXII. Constraints on inflation,''
  arXiv:1303.5082 [astro-ph.CO].


\bibitem{Ade:2013nlj}
  P.~A.~R.~Ade {\it et al.}  [Planck Collaboration],
  ``Planck 2013 results. XXIII. Isotropy and Statistics of the CMB,''
  arXiv:1303.5083 [astro-ph.CO].

\bibitem{Eriksen:2003db}
  H.~K.~Eriksen, F.~K.~Hansen, A.~J.~Banday, K.~M.~Gorski and P.~B.~Lilje,
  ``Asymmetries in the Cosmic Microwave Background anisotropy field,''
  Astrophys.\ J.\  {\bf 605} (2004) 14
   [Erratum-ibid.\  {\bf 609} (2004) 1198]
  [astro-ph/0307507];
  H.~K.~Eriksen, A.~J.~Banday, K.~M.~Gorski, F.~K.~Hansen and P.~B.~Lilje,
  ``Hemispherical power asymmetry in the three-year Wilkinson Microwave Anisotropy Probe sky maps,''
  Astrophys.\ J.\  {\bf 660}, L81 (2007)
  [astro-ph/0701089].

\bibitem{Bennett:2010jb}
  C.~L.~Bennett, R.~S.~Hill, G.~Hinshaw, D.~Larson, K.~M.~Smith, J.~Dunkley, B.~Gold and M.~Halpern {\it et al.},
  ``Seven-Year Wilkinson Microwave Anisotropy Probe (WMAP) Observations: Are There Cosmic Microwave Background Anomalies?,''
  Astrophys.\ J.\ Suppl.\  {\bf 192}, 17 (2011)
  [arXiv:1001.4758 [astro-ph.CO]].

\bibitem{Hanson:2010gu}
  D.~Hanson, A.~Lewis and A.~Challinor,
  ``Asymmetric Beams and CMB Statistical Anisotropy,''
  Phys.\ Rev.\ D {\bf 81}, 103003 (2010)
  [arXiv:1003.0198 [astro-ph.CO]].



\bibitem{Dai:2013kfa}
  L.~Dai, D.~Jeong, M.~Kamionkowski and J.~Chluba,
  ``The Pesky Power Asymmetry,''
  arXiv:1303.6949 [astro-ph.CO].





\bibitem{Pullen:2007tu}
  A.~R.~Pullen and M.~Kamionkowski,
  ``Cosmic Microwave Background Statistics for a Direction-Dependent Primordial Power Spectrum,''
  Phys.\ Rev.\ D {\bf 76}, 103529 (2007)
  [arXiv:0709.1144 [astro-ph]].

  \bibitem{Grishchuk1978}
  L.P. Grishchuk and I.B. Zel'dovich, Soviet Astronomy {\bf{22}}, 125 (1978).


\bibitem{Erickcek:2008sm}
  A.~L.~Erickcek, M.~Kamionkowski and S.~M.~Carroll,
  ``A Hemispherical Power Asymmetry from Inflation,''
  Phys.\ Rev.\ D {\bf 78}, 123520 (2008)
  [arXiv:0806.0377 [astro-ph]].


\bibitem{Gordon:2006ag}
  C.~Gordon,
  ``Broken Isotropy from a Linear Modulation of the Primordial Perturbations,''
  Astrophys.\ J.\  {\bf 656}, 636 (2007)
  [astro-ph/0607423].





\bibitem{Sheth:1999mn}
  R.~K.~Sheth and G.~Tormen,
  ``Large scale bias and the peak background split,''
  Mon.\ Not.\ Roy.\ Astron.\ Soc.\  {\bf 308}, 119 (1999)
  [astro-ph/9901122].



\bibitem{Schmidt:2012ky}
  F.~Schmidt and L.~Hui,
  ``Cosmic Microwave Background Power Asymmetry from Non-Gaussian Modulation,''
  Phys.\ Rev.\ Lett.\  {\bf 110}, 011301 (2013)
  [Publisher-note {\bf 110}, 059902 (2013)]
  [arXiv:1210.2965 [astro-ph.CO]].


\bibitem{Prunet:2004zy}
  S.~Prunet, J.~-P.~Uzan, F.~Bernardeau and T.~Brunier,
  ``Constraints on mode couplings and modulation of the CMB with WMAP data,''
  Phys.\ Rev.\ D {\bf 71}, 083508 (2005)
  [astro-ph/0406364].

\bibitem{Byrnes:2011ri} 
  C.~T.~Byrnes, S.~Nurmi, G.~Tasinato and D.~Wands,
  ``Inhomogeneous non-Gaussianity,''
  JCAP {\bf 1203}, 012 (2012)
  [arXiv:1111.2721 [astro-ph.CO]].



\bibitem{Erickcek:2008jp}
  A.~L.~Erickcek, S.~M.~Carroll and M.~Kamionkowski,
  ``Superhorizon Perturbations and the Cosmic Microwave Background,''
  Phys.\ Rev.\ D {\bf 78}, 083012 (2008)
  [arXiv:0808.1570 [astro-ph]].




\bibitem{Turner:1991dn}
  M.~S.~Turner,
  ``A Tilted universe (and other remnants of the preinflationary universe),''
  Phys.\ Rev.\ D {\bf 44}, 3737 (1991).


\bibitem{Zibin:2008fe}
  J.~P.~Zibin and D.~Scott,
  ``Gauging the cosmic microwave background,''
  Phys.\ Rev.\ D {\bf 78}, 123529 (2008)
  [arXiv:0808.2047 [astro-ph]].



\bibitem{Sachs:1967er}
  R.~K.~Sachs and A.~M.~Wolfe,
  ``Perturbations of a cosmological model and angular variations of the microwave background,''
  Astrophys.\ J.\  {\bf 147}, 73 (1967)
  [Gen.\ Rel.\ Grav.\  {\bf 39}, 1929 (2007)].


\bibitem{Bruni:1993dx}
  M.~Bruni and D.~H.~Lyth,
  ``Peculiar velocity, cosmic perturbation theory and the CMB anisotropy,''
  Phys.\ Lett.\ B {\bf 323}, 118 (1994)
  [astro-ph/9307036].



\bibitem{Castro:2003bk}
  P.~G.~Castro, M.~Douspis and P.~G.~Ferreira,
  ``Scale of homogeneity of the universe from WMAP,''
  Phys.\ Rev.\ D {\bf 68}, 127301 (2003)
  [astro-ph/0309320].



\bibitem{Maldacena:2002vr}
  J.~M.~Maldacena,
  ``Non-Gaussian features of primordial fluctuations in single field inflationary models,''
  JHEP {\bf 0305}, 013 (2003)
  [astro-ph/0210603].

\bibitem{Kundu:2011sg} 
  S.~Kundu,
  ``Inflation with General Initial Conditions for Scalar Perturbations,''
  JCAP {\bf 1202}, 005 (2012)
  [arXiv:1110.4688 [astro-ph.CO]].

\bibitem{Lyth:2013vha}
  D.~H.~Lyth,
  ``The CMB asymmetry from inflation,''
  arXiv:1304.1270 [astro-ph.CO].



\bibitem{Liu:2013kea} 
  Z.~-G.~Liu, Z.~-K.~Guo and Y.~-S.~Piao,
  ``Obtaining the CMB anomalies with a bounce from the contracting phase to inflation,''
  arXiv:1304.6527 [astro-ph.CO].



\bibitem{Hirata:2009ar}
  C.~M.~Hirata,
  ``Constraints on cosmic hemispherical power anomalies from quasars,''
  JCAP {\bf 0909}, 011 (2009)
  [arXiv:0907.0703 [astro-ph.CO]].



\bibitem{Gordon:2005ai}
  C.~Gordon, W.~Hu, D.~Huterer and T.~M.~Crawford,
  ``Spontaneous isotropy breaking: a mechanism for cmb multipole alignments,''
  Phys.\ Rev.\ D {\bf 72}, 103002 (2005)
  [astro-ph/0509301].
 ;
  J.~F.~Donoghue, K.~Dutta and A.~Ross,
  ``Non-isotropy in the CMB power spectrum in single field inflation,''
  Phys.\ Rev.\ D {\bf 80}, 023526 (2009)
  [astro-ph/0703455 [ASTRO-PH]].


\bibitem{Wang:2013lda}
  L.~Wang and A.~Mazumdar,
  ``Small non-Gaussianity and dipole asymmetry in the CMB,''
  arXiv:1304.6399 [astro-ph.CO].




\bibitem{Lyth:2001nq}
  D.~H.~Lyth and D.~Wands,
  ``Generating the curvature perturbation without an inflaton,''
  Phys.\ Lett.\ B {\bf 524}, 5 (2002)
  [hep-ph/0110002].

\bibitem{Linde:1996gt} 
  A.~D.~Linde and V.~F.~Mukhanov,
  ``Nongaussian isocurvature perturbations from inflation,''
  Phys.\ Rev.\ D {\bf 56}, 535 (1997)
  [astro-ph/9610219].


\bibitem{Namjoo:2012aa}
  M.~H.~Namjoo, H.~Firouzjahi and M.~Sasaki,
  ``Violation of non-Gaussianity consistency relation in a single field inflationary model,''
  arXiv:1210.3692 [astro-ph.CO].

\bibitem{Chen:2013aj}
  X.~Chen, H.~Firouzjahi, M.~H.~Namjoo and M.~Sasaki,
  ``A Single Field Inflation Model with Large Local Non-Gaussianity,''
  arXiv:1301.5699 [hep-th].

\bibitem{RenauxPetel:2010ty}
  S.~Renaux-Petel,
  ``On the squeezed limit of the bispectrum in general single field inflation,''
  JCAP {\bf 1010}, 020 (2010)
  [arXiv:1008.0260 [astro-ph.CO]].


\bibitem{Creminelli:2004yq}
  P.~Creminelli and M.~Zaldarriaga,
  ``Single field consistency relation for the 3-point function,''
  JCAP {\bf 0410}, 006 (2004)
  [astro-ph/0407059].


%



\bibitem{Ganc:2010ff}
  J.~Ganc and E.~Komatsu,
  ``A new method for calculating the primordial bispectrum in the squeezed limit,''
  JCAP {\bf 1012}, 009 (2010)
  [arXiv:1006.5457 [astro-ph.CO]].


\bibitem{Efstathiou:2003tv}
  G.~Efstathiou,
  ``A Maximum likelihood analysis of the low CMB multipoles from WMAP,''
  Mon.\ Not.\ Roy.\ Astron.\ Soc.\  {\bf 348}, 885 (2004)
  [astro-ph/0310207].

\bibitem{Lyth:2007jh}
  D.~H.~Lyth,
  JCAP {\bf 0712}, 016 (2007)
  [arXiv:0707.0361 [astro-ph]].

\bibitem{Ade:2013ydc}
  P.~A.~R.~Ade {\it et al.}  [Planck Collaboration],
  ``Planck 2013 Results. XXIV. Constraints on primordial non-Gaussianity,''
  arXiv:1303.5084 [astro-ph.CO].

\bibitem{Sasaki:1995aw}
  M.~Sasaki, E.~D.~Stewart,
  ``A General analytic formula for the spectral index of
 the density perturbations produced during inflation,''
  Prog.\ Theor.\ Phys.\  {\bf 95}, 71-78 (1996).
  [astro-ph/9507001].


\bibitem{Wands:2000dp}
  D.~Wands, K.~A.~Malik, D.~H.~Lyth {\it et al.},
  ``A New approach to the evolution of cosmological perturbations on large scales,''
  Phys.\ Rev.\  {\bf D62}, 043527 (2000).
  [astro-ph/0003278].




\bibitem{Erickcek:2009at}
  A.~L.~Erickcek, C.~M.~Hirata and M.~Kamionkowski,
  ``A Scale-Dependent Power Asymmetry from Isocurvature Perturbations,''
  Phys.\ Rev.\ D {\bf 80}, 083507 (2009)
  [arXiv:0907.0705 [astro-ph.CO]].


\bibitem{McDonald:2013aca} 
  J.~McDonald,
  ``Isocurvature and Curvaton Perturbations with Red Power Spectrum and Large Hemispherical Asymmetry,''
  arXiv:1305.0525 [hep-ph].

\bibitem{Assadullahi:2013ey}
  H.~Assadullahi, H.~Firouzjahi, M.~H.~Namjoo and D.~Wands,
  ``Modulated curvaton decay,''
  JCAP {\bf 1303}, 041 (2013)
  [arXiv:1301.3439 [hep-th]].


\bibitem{Fonseca:2012cj}
  J.~Fonseca and D.~Wands,
  ``Primordial non-Gaussianity from mixed inflaton-curvaton perturbations,''
  JCAP {\bf 1206}, 028 (2012)
  [arXiv:1204.3443 [astro-ph.CO]].



\bibitem{Langlois:2013dh}
  D.~Langlois and T.~Takahashi,
  ``Density Perturbations from Modulated Decay of the Curvaton,''
  JCAP {\bf 1304}, 014 (2013)
  [arXiv:1301.3319 [astro-ph.CO]].








\bibitem{Bernardeau:2004zz}
  F.~Bernardeau, L.~Kofman and J.~-P.~Uzan,
  ``Modulated fluctuations from hybrid inflation,''
  Phys.\ Rev.\ D {\bf 70}, 083004 (2004)
  [astro-ph/0403315].


\bibitem{Dvali:2003em}
  G.~Dvali, A.~Gruzinov and M.~Zaldarriaga,
  ``A new mechanism for generating density perturbations from inflation,''
  Phys.\ Rev.\ D {\bf 69}, 023505 (2004)
  [astro-ph/0303591].

\bibitem{Lyth:2005qk}
  D.~H.~Lyth,
  ``Generating the curvature perturbation at the end of inflation,''
  JCAP {\bf 0511}, 006 (2005).
  [astro-ph/0510443].


\bibitem{Assadullahi:2012yi}
  H.~Assadullahi, H.~Firouzjahi, M.~H.~Namjoo and D.~Wands,
  JCAP {\bf 1212}, 024 (2012)
  [arXiv:1207.7006 [astro-ph.CO]].










\bibitem{Copeland:1994vg}
 A.~R.~Liddle, D.~H.~Lyth, E.~D.~Stewart and D.~Wands,
  ``False vacuum inflation with Einstein gravity,''
  Phys.\ Rev.\  D {\bf 49}, 6410 (1994)
  [arXiv:astro-ph/9401011].

\bibitem{Linde:1993cn}
  A.~D.~Linde,
  ``Hybrid inflation,''
  Phys.\ Rev.\  D {\bf 49}, 748 (1994)
  [arXiv:astro-ph/9307002].


\bibitem{Abolhasani:2010kr}
  A.~A.~Abolhasani, H.~Firouzjahi,
  ``No Large Scale Curvature Perturbations during Waterfall of Hybrid Inflation,''
  Phys.\ Rev.\  {\bf D83}, 063513 (2011).
  [arXiv:1005.2934 [hep-th]].









\bibitem{Lyth:2010ch}
  D.~H.~Lyth,
  ``Issues concerning the waterfall of hybrid inflation,''
  Prog.\ Theor.\ Phys.\ Suppl.\  {\bf 190}, 107 (2011)
  [arXiv:1005.2461 [astro-ph.CO]].



\bibitem{Fonseca:2010nk}
  J.~Fonseca, M.~Sasaki, D.~Wands,
``Large-scale Perturbations from the Waterfall Field in Hybrid Inflation,''
  JCAP {\bf 1009}, 012 (2010).
  [arXiv:1005.4053 [astro-ph.CO]].

\bibitem{Abolhasani:2011yp}
  A.~A.~Abolhasani, H.~Firouzjahi and M.~Sasaki,
  ``Curvature perturbation and waterfall dynamics in hybrid inflation,''
  JCAP {\bf 1110}, 015 (2011)
  [arXiv:1106.6315 [astro-ph.CO]].

\bibitem{Gong:2010zf}
  J.~-O.~Gong, M.~Sasaki,
  ``Waterfall field in hybrid inflation and curvature perturbation,''
  JCAP {\bf 1103}, 028 (2011).
  [arXiv:1010.3405 [astro-ph.CO]].



\bibitem{Clesse:2010iz}
  S.~Clesse,
  ``Hybrid inflation along waterfall trajectories,''
  Phys.\ Rev.\ D {\bf 83}, 063518 (2011)
  [arXiv:1006.4522 [gr-qc]].

\bibitem{Martin:2011ib}
  J.~Martin and V.~Vennin,
  ``Stochastic Effects in Hybrid Inflation,''
  Phys.\ Rev.\ D {\bf 85}, 043525 (2012)
  [arXiv:1110.2070 [astro-ph.CO]].

\bibitem{Mulryne:2011ni}
  D.~Mulryne, S.~Orani and A.~Rajantie,
  ``Non-Gaussianity from the hybrid potential,''
  Phys.\ Rev.\ D {\bf 84}, 123527 (2011)
  [arXiv:1107.4739 [hep-th]].

















\end{thebibliography}
\end{document}